\documentclass{article}
\usepackage[margin=1in]{geometry}
\usepackage[utf8]{inputenc}
\usepackage[numbers,sort&compress]{natbib}

\usepackage{makecell,booktabs}
\usepackage{float}

\usepackage{placeins}
\renewcommand{\bibsection}{\subsubsection*{\bibname}}

\usepackage[dvipsnames,table,xcdraw]{xcolor}

\usepackage{comment}
\usepackage{pdfcomment}
\usepackage{environ}
\RenewEnviron{comment}{\pdfcomment{\BODY}}

\usepackage{dcolumn}
\usepackage{booktabs}
\usepackage{tikz}
\usetikzlibrary{positioning,shapes,arrows}
\usetikzlibrary{decorations.pathreplacing}

\usepackage{graphicx}

\usepackage{subcaption}

\usepackage{algorithm}
\usepackage[noend]{algpseudocode}

\usepackage{amsmath}
\usepackage{amssymb}

\algrenewcommand\algorithmicindent{1em}

\usepackage{amsthm}

\theoremstyle{definition}

\theoremstyle{remark}

\makeatletter
\def\thm@space@setup{%
  \thm@preskip=0mm plus 0.5mm minus 0mm
  \thm@postskip=0mm plus 0.5mm minus 0mm
}
\makeatother
\usepackage{enumerate}

\usepackage{wrapfig}
\usepackage{amssymb}

\title{Mathematical construction of a low-bias high-resolution deprivation index for the United States}
\author{Amin Ghafourian$^{1,2,\ast}$, Noli Brazil$^3$, and Thilo Gross$^{1,4,5,6}$}
\date{\today}
\begin{document}
\maketitle
\vspace{-0.5cm}
\begin{center}
\begin{minipage}{.75\linewidth}
    \footnotesize
    $^1$University of California, Davis, Department of Computer Science, USA\\
    $^2$University of California, Davis, Department of Mechanical and Aerospace Engineering, USA\\
    $^3$University of California, Davis, Department of Human Ecology, USA\\
    $^4$HIFMB, Helmholtz Institute for Functional Marine Biodiversity, Germany\\
    $^5$Carl-Von-Ossietzky University, Germany\\
    $^6$Alfred-Wegener Institute, Helmholtz Centre for Marine and Polar Research, Germany\\
    $^\ast$corresponding author: aghafourian@ucdavis.edu 
\end{minipage}
\end{center}

\begin{abstract}
The construction of deprivation indices is complicated by the inherent ambiguity in defining deprivation as well as the potential for partisan manipulation. Nevertheless, deprivation indices provide an essential tool for mitigating the effects of deprivation and reducing it through policy interventions. Here we demonstrate the construction of a deprivation index using diffusion maps, a manifold learning technique capable of finding the variables that optimally describe the variations in a dataset in the sense of preserving pairwise relationships among the data points. The method is applied to the 2010 US decennial census. In contrast to other methods the proposed procedure does not select particular columns from the census, but rather constructs an indicator of deprivation from the complete dataset. Due to its construction the proposed index does not introduce biases except those already present in the source data, does not require normative judgment regarding the desirability of certain life styles, and is highly resilient against attempts of partisan manipulation. We demonstrate that the new index aligns well with established income-based deprivation indices but deviates in aspects that are perceived as problematic in some of the existing indices. The proposed procedure provides an efficient way for constructing accurate, high resolution indices. These indices can thus have the potential to become powerful tools for the academic study of social structure as well as political decision making. 
\end{abstract}

\section*{Introduction}
Over the past decades rising productivity and international cooperation have led to a rapid growth of wealth throughout the developed world. Nevertheless, the degree to which individuals profit from this growth is increasingly uneven and thus a significant proportion of the population find themselves in worse economic circumstances, absolutely or relatively, than they would have been in previous decades \cite{piketty2015}. Simultaneously also the disparity between favored and disadvantaged neighborhoods is increasing \cite{florida2017}. Mounting evidence indicates that the social, physical and economic characteristics of residential environments impact social mobility \cite{chetty2014, sampson2012,galster2019}. Even in the United States, where social mobility is counted among the nation's foundational values, people from poor backgrounds find themselves locked in social environments that further disenfranchise them \cite{sharkey2012}. 

The rise in spatial inequality, the inequality across places, and the backlash it has engendered in the form of populism has convinced many policymakers of the need to embrace place-based policies to bolster the conditions of declining communities \cite{neumark2015place}. For this purpose areas in need of social and economic investment are sometimes still identified based on a single statistical variable, such as income level \cite{us2019census,jargowsky1997poverty,berube2007geography}. However, deprivation is an inherently multidimensional construct, where well-being is both impacted and reflected by a multitude of factors.

The acknowledgment that deprivation is multidimensional has led to the adoption of deprivation indices that take multiple factors into account. For example in the UK the so-called indices of multiple deprivation factor several well-being domains including crime and safety, housing, education, and employment \cite{payne2012uk}. Although the United States has yet to adopt a national deprivation index, local public agencies have constructed their own indices in order to efficiently allocate resources to the most deprived areas \cite{us2019census,united2019hhs,fox2018supplemental,bohn2013california,renwick2011geographic}.

The construction of a national-level multi-factorial deprivation index is difficult due to issues related to the variable selection method and the statistical model used to construct the index. A complex index designed by experts is highly susceptible to accusations of partisan bias. Furthermore, current approaches make judgments regarding which contextual characteristics are relevant for adequate functioning and a minimally acceptable standard for each of these characteristics. Moreover, proxies used to assess deprivation remain varied and unstandardized. These approaches may also overlook the geographic heterogeneity of the US population or the ways in which uniquely characterized neighborhoods, such as retirement communities or college campuses, complicate a unidimensional application of deprivation. This defines a need for a methodology that allows for the construction of a deprivation index that minimizes biases and is robust against intentional manipulation.   

An analysis of the UK census showed that economic deprivation is one of the two most important variables that shaped census responses \cite{barter2019manifold}. Importantly that study did not look for deprivation specifically but used a so-called diffusion map, a general mathematical methodology, to identify explanatory variables in large data sets. Economic deprivation emerged as an explanatory variable although the UK census does not contain income information, instead the method detected a similarity in living conditions in subset of the population that was reflected in several hundred different statistics reported in the census.     

In this paper we use diffusion maps to construct a social deprivation index for the United States. The diffusion map is a nonlinear method that is applicable to large and complex datasets. Nevertheless it is mathematically simple and builds on a strong physical intuition \cite{coifman2005geometric,coifman2006diffusion,lafon2004diffusion,lafon2006data}. It contains only few tunable parameters governed by strong rationals and results are typically robust against parameter variation. Hence the diffusion map leaves almost no room for partisan manipulation and does not introduce biases beyond those that may already exist in the source data.

Our analysis produces an index that aligns well with previous deprivation indices but offers a more comprehensive and detailed picture. This picture reveals large disadvantaged regions, but also highlights very high heterogeneity at the local scale. While diffusion maps of the UK cities revealed the divide between poor and middle class, it is very affluent areas that stand out most in the US. One interpretation of this result is that unlike deprivation in the UK, where the middle class neighborhoods distance themselves from poor areas, it is the affluent neighborhoods in the United States that distance themselves from the middle class.

\section*{Census Analysis with Diffusion Maps}
We start our analysis not by trying to identify the deprived areas, but rather by asking which are the most similar, where the notions of similarity used are mathematically discovered from the dataset itself. The analysis considers census tract level data from the United States Census \cite{uscensustechnical,us2010census}. Census tracts are administrative units that have been used as proxies for neighborhoods in community and neighborhood level analysis and are designed to represent social and economically homogeneous groups of approximately 1200 to 8000 persons. For our main analysis we removed information on race. Though not strictly necessary for the method to work, this was done as a precaution to avoid racial bias in the results. The result is a dataset containing $1385$ parameters for each of the $73057$ census tracts in the US. 

We then computed the similarity between all pairs of tracts using a common metric of similarity, the inverse of the Euclidean distance between the census data vectors. This metric of similarity is well-suited for comparing very similar tracts, but performs poorly when considering dissimilar areas \cite{lafon2004diffusion,barter2019manifold}. 

To avoid accumulation of error from such dissimilar comparisons, we threshold the initial similarity measurements. This is done, by checking the similarity between each pair of nodes $a,b$ and setting the similarity to zero if it is not among the 9 greatest values of similarity that either $a$ or $b$ have with any other node. The matrix of pairwise similarities can be interpreted as a weighted adjacency matrix of a sparse network. In this network the nodes are census tracts and the links between tracts indicate similarity. 

We then constructed more refined notions of similarity by computing the normalized Laplacian matrix corresponding to the network and computing its eigenvectors (see Appendix). For a high-dimensional dataset there are many such eigenvectors which represent different alternative notions of similarity. Every eigenvector assigns a score to each of the census tracts which implies an ordering of the tracts. Thus the eigenvectors identify new emergent variables that describe the dataset. For example, below we argue the values of one of the eigenvectors acts as an indicator of deprivation by ordering the census tracts from the most deprived to least deprived. 

Each eigenvector is associated with an eigenvalue. Eigenvalues indicate the relative importance of the differences picked up by the eigenvectors. In the implementation of the diffusion map described here the most important notions of similarity are those that correspond to eigenvalues close to zero. 

The steps above described the diffusion map \cite{lafon2004diffusion}, a straightforward mathematical procedure, where the only user defined parameters are the kernel used for the construction of initial similarities and the choice of threshold value. Both of these are guided by physical principles and cannot plausibly be used to manufacture ideologically desirable results (see Appendix). By contrast the interpretation (e.g.~``deprivation'') that we attribute to an eigenvector requires human intuition, and hence must be verified using additional data or analysis. 

One problem that had to be overcome in our analysis is that for certain tracts in southern states only partial census statistics were reported. However, previous work \cite{barter2019manifold} has shown that such gaps can be closed by running the analysis without these incomplete tracts and then assigning them the vector elements from the most similar analyzed tract.

For illustration we visualize  the first four census eigenvectors in an area around Los Angeles (Fig.~1). The most important eigenvector reveals a sharply localized pattern, where some tracts receive large magnitude entries whereas the entries in most others are virtually zero. This shows that there is a set of tracts where census responses are similar, while being different from the responses in most other tracts. To interpret this result we manually identified the respective tracts and found they coincided with universities.  

\begin{figure*}[tbhp]
\centering
\includegraphics[width=\textwidth]{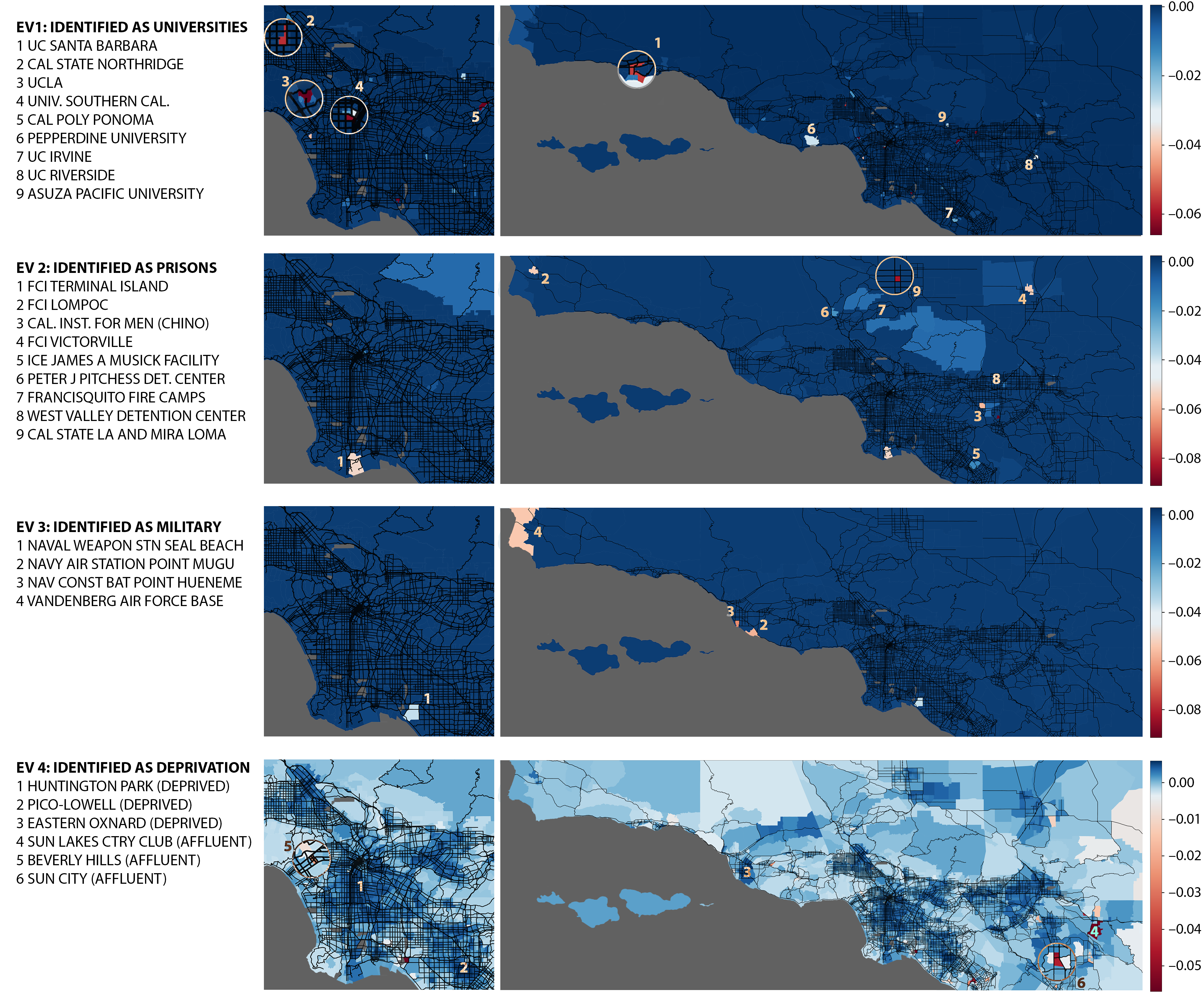}
\caption{Patterns of social similarity in Los Angeles (center) and the surrounding area (right). The first eigenvector is found to highlight colleges. The second eigenvector detects prisons and detention centers. Military bases and training facilities are highlighted in the third eigenvector. The fourth eigenvector provides an accurate high-resolution proxy for deprivation. The color corresponds directly to eigenvector entries (arbitrary units). Grey indicates uninhabited areas. Places of interest are marked with numbers. Some of these places where zoomed (circles) to make small features visible.
}
\end{figure*}

Similarly, the second and third largest eigenvectors exhibit localized patterns corresponding to the locations of correctional facilities and military bases. Collectively, these first three eigenvectors highlight strong social differences, which exist for clear extrinsic reasons, i.e.~enrollment in college, military service, or incarceration.

\begin{figure*}[tbhp]
\centering
\begin{subfigure}{\textwidth}
  \centering
  \includegraphics[width=0.95\textwidth]{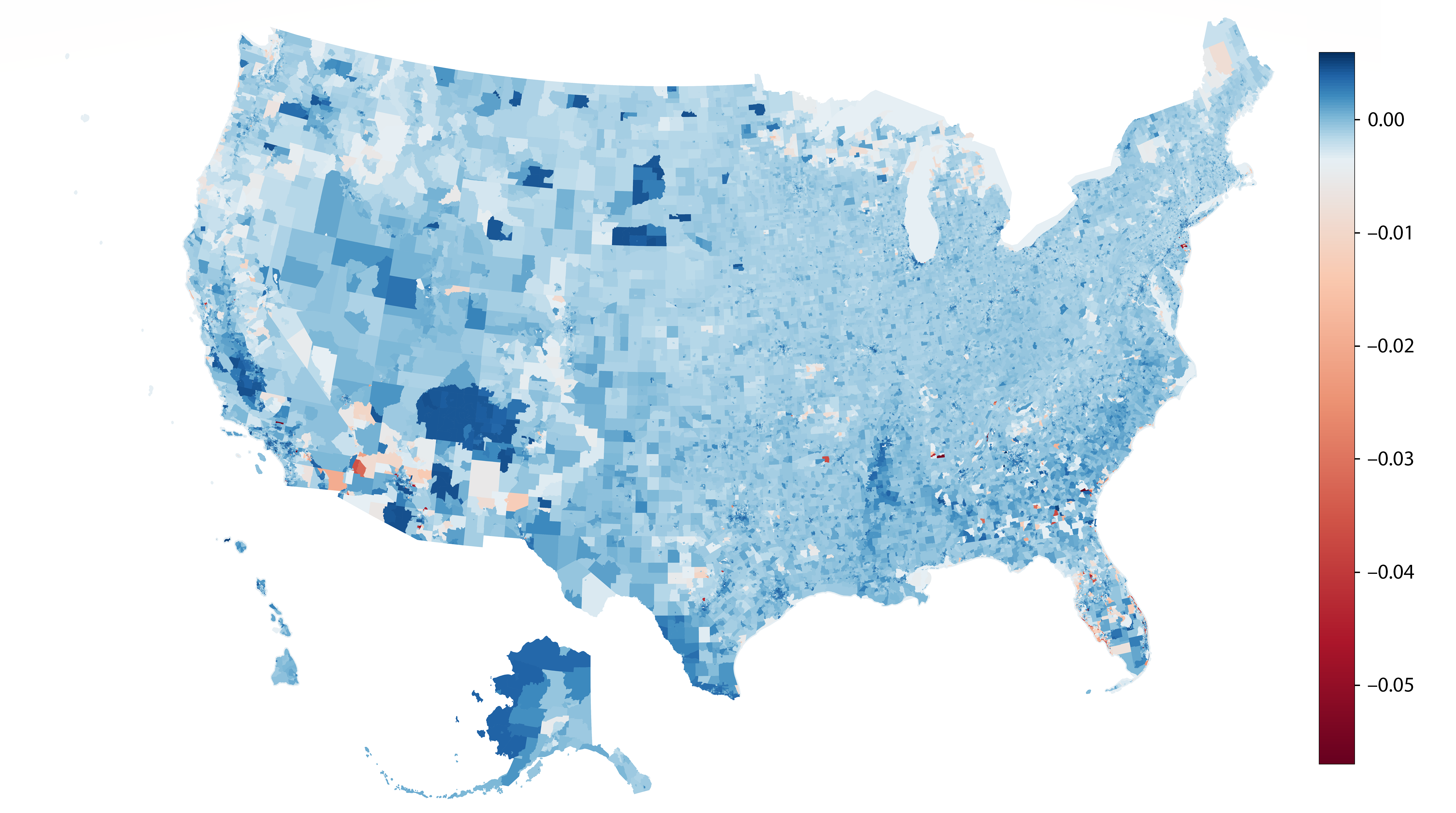}
    \caption{}
\end{subfigure}
\begin{subfigure}{.3\textwidth}
  \centering
  \includegraphics[height=4.7cm]{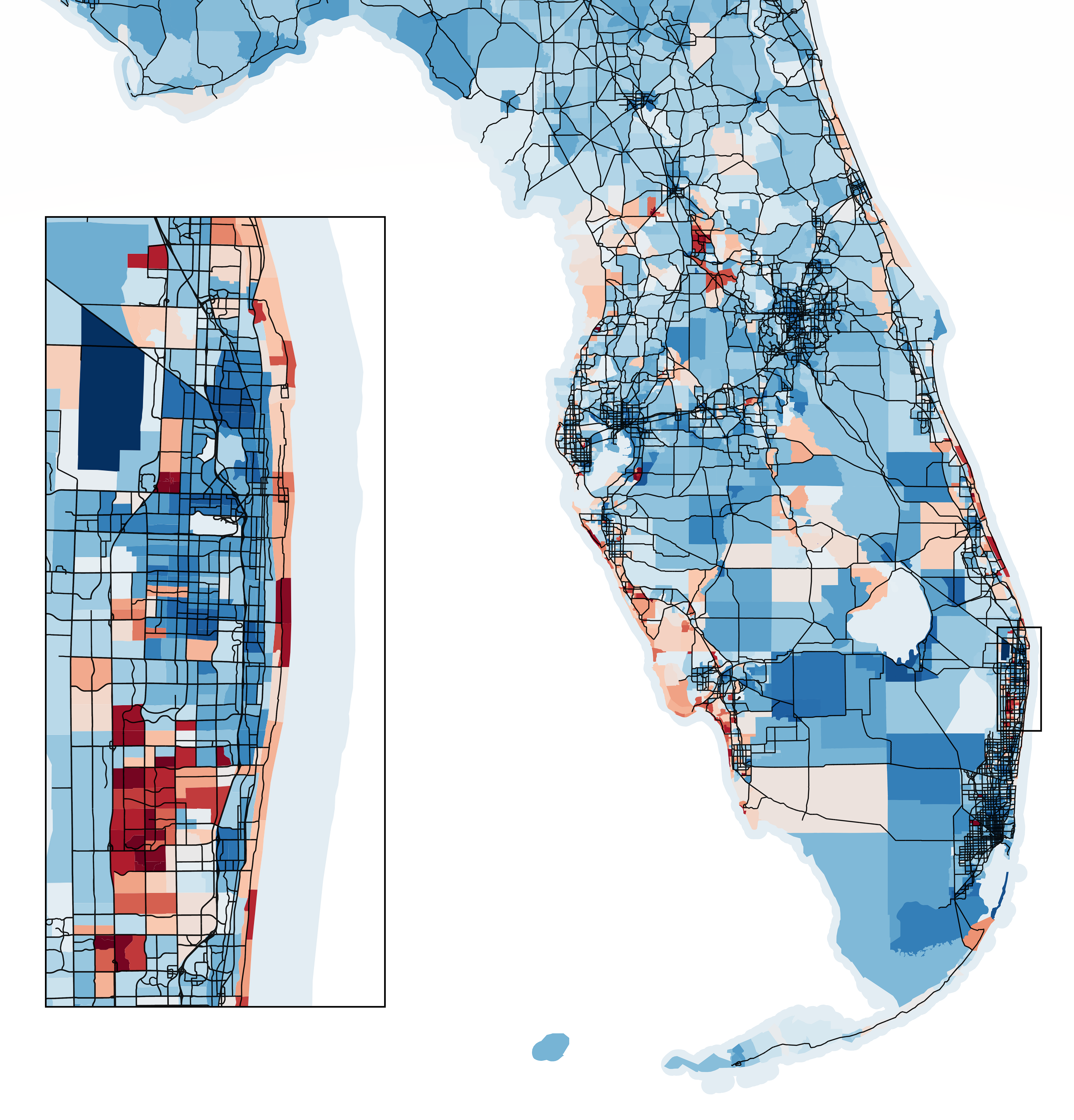}
    \caption{}
\end{subfigure}%
\begin{subfigure}{.65\textwidth}
  \centering
  \includegraphics[height=4.7cm]{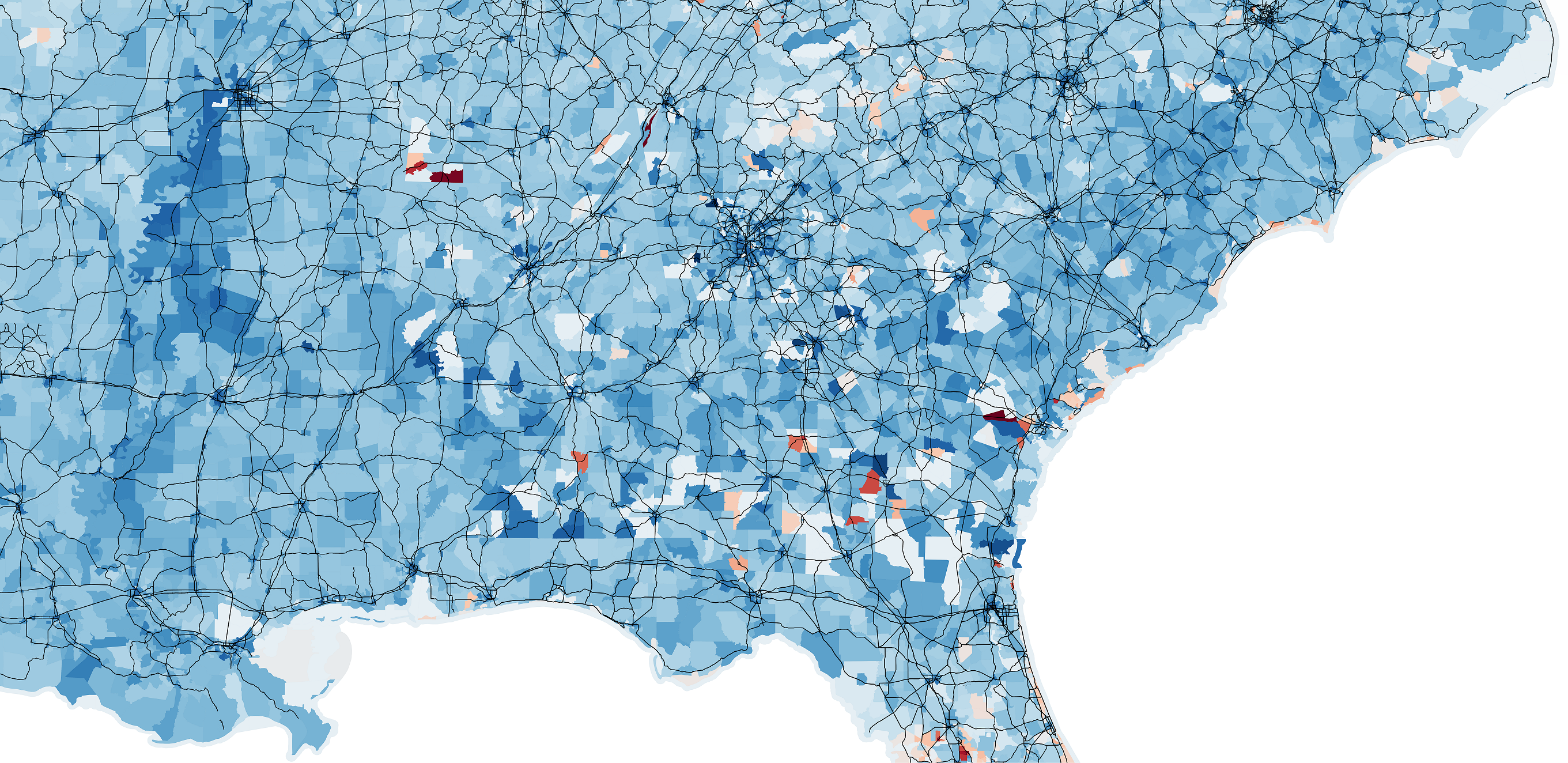}
    \caption{}
\end{subfigure}
\begin{subfigure}{.42\textwidth}
  \centering
  \includegraphics[height=5.9cm]{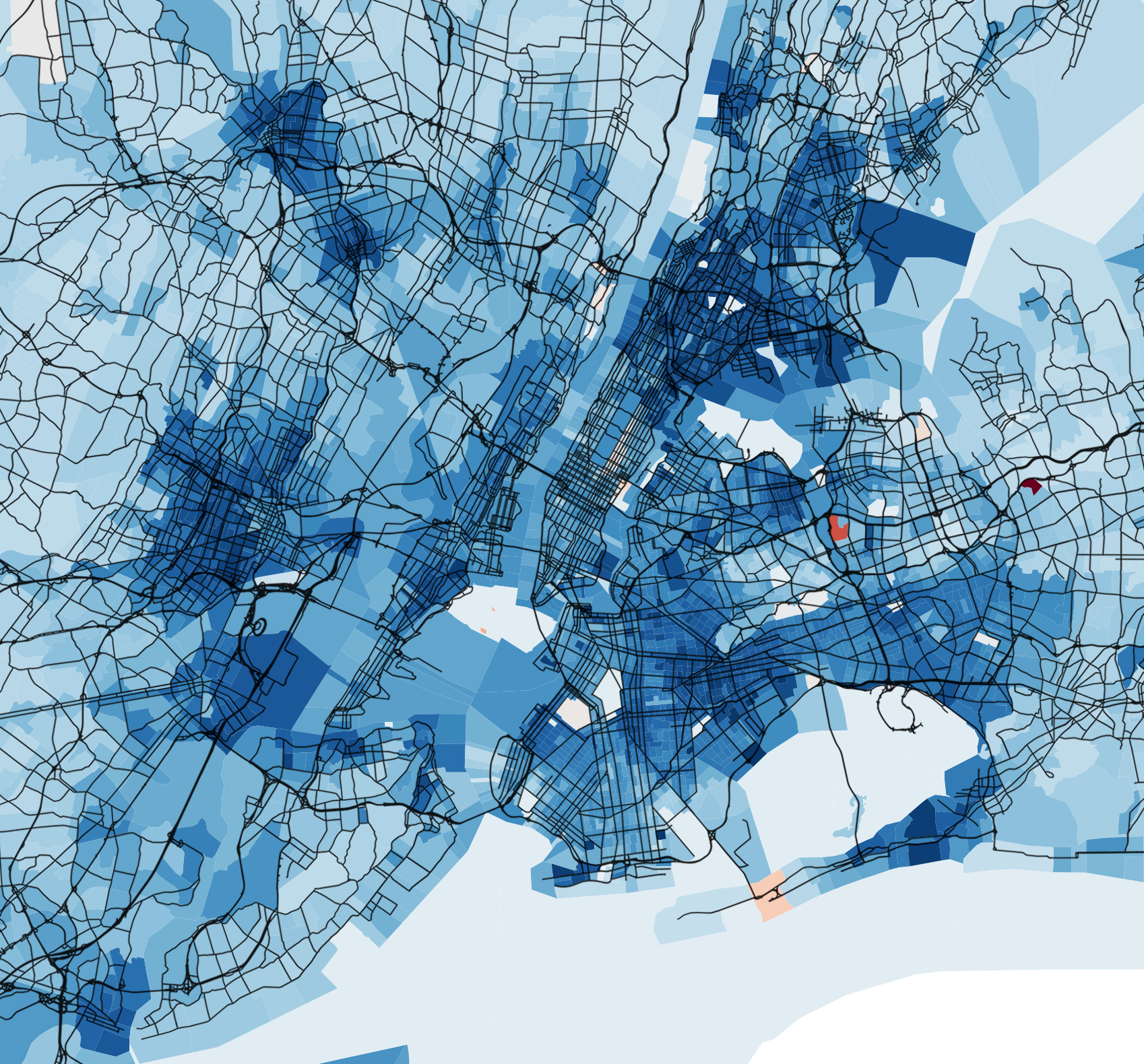}
    \caption{}
\end{subfigure}%
\begin{subfigure}{.53\textwidth}
  \centering
  \includegraphics[height=5.9cm]{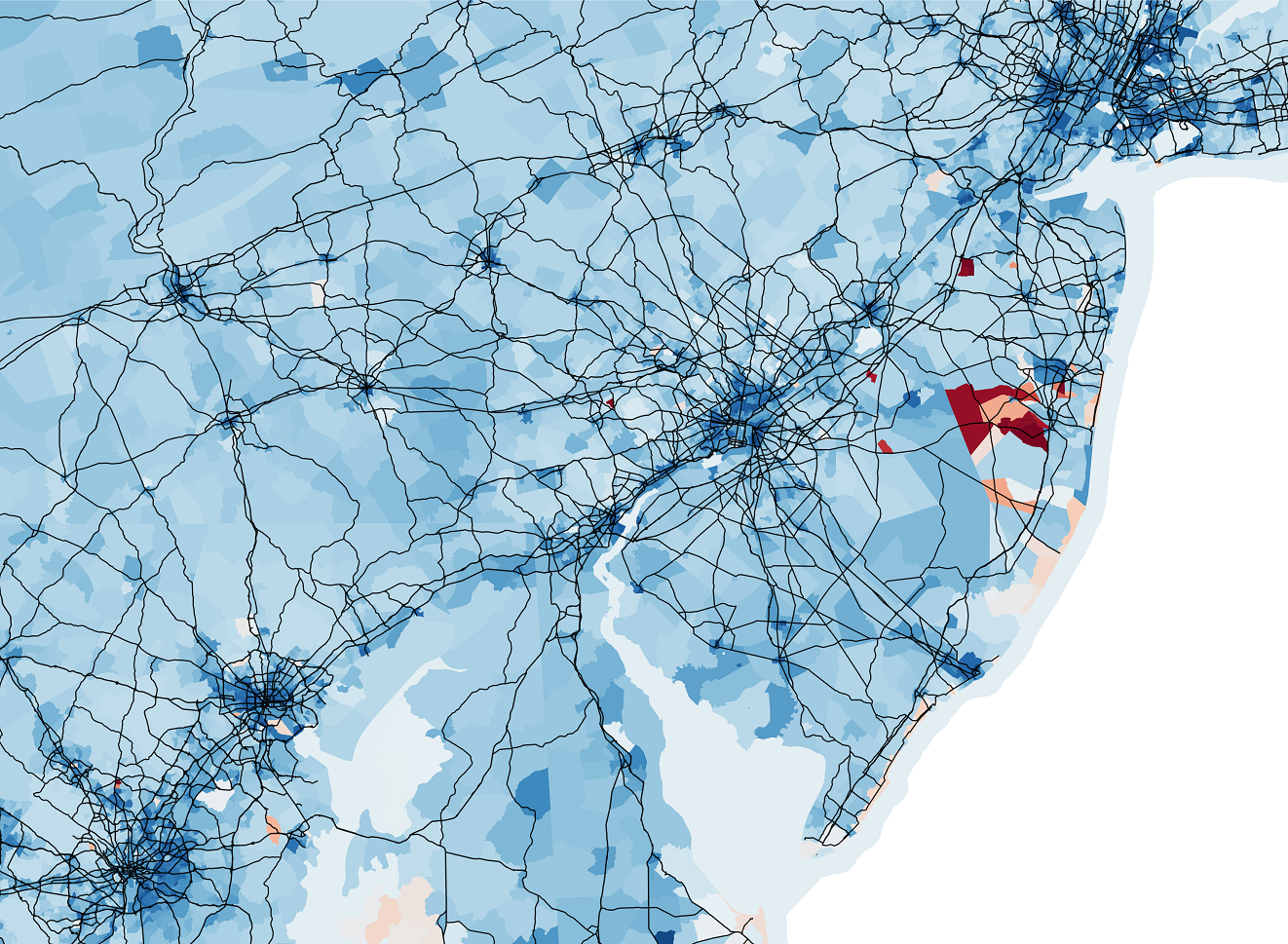}
    \caption{}
\end{subfigure}
\caption{Diffusion-map-based deprivation index in the US shown for the 50 states and Washington, D.C. (a), parts of Florida (b), south (c), New York City (d), and east and northeast (e). Darker shade of blue indicates higher deprivation. Areas with zero population are shown in white.
}
\end{figure*}

\section*{Deprivation in the USA}
We hypothezise that the fourth eigenvector is an indicator of deprivation. By visual inspection one can confirm that tracts that are assigned negative numbers are affluent areas, retirement homes and country clubs, whereas tracts assigned positive numbers coincide with deprived areas. To test the hypothesis further we correlated the results in the Los Angeles county with an income-based deprivation index that exists for this area \cite{california2017calenviroscreen,calenviro2017OEHHA}. We find a rank correlation of $0.84$, which strongly supports the interpretation of the eigenvector as a deprivation indicator. 

We found that for this purpose the diffusion map performs better than comparable methods (e.g.~PCA yields a rank correlation of only $0.58$ with the LA index, see Appendix).

We note that neither our eigenvector nor other, e.g.~income-based indicators reflect a ground truth. Particularly we see the use of income-based indicators on a national level as problematic due to the different rent and consumer price levels. By contrast, differences in living circumstances reflected in many census variables and picked up by the eigenvector may reveal a more precise picture.

We show some of the attributes of the top and bottom 10 census tracts in the diffusion map index (Table 1) from an independent dataset \cite{united2013american, subject2012census}. The statistics on income, percentage of population below poverty level, and housing costs that are significantly affected by the location vary within each group and do not delineate a clear separation margin between the two extremes; On the other hand, statistics such as the educational attainment and percentage employment that are less dependent on circumstances like location remain consistent within each group and widely differ across the two extremes. Overall, among the tracts highlighted as deprived we see several different forms of deprivation, ranging from tracts with incredibly low housing costs to comparatively expensive but contextually poor inner city neighborhoods. This diversity shows that the diffusion map has successfully identified a commonality that underlies the diverse forms of deprivation.

\begin{table*}
\footnotesize
\centering
\caption{Socioeconomic statistics for the 10 most and least deprived census tracts in the diffusion map deprivation index (respectively top 10 and bottom 10 entries) taken from the 2008--2012 American Community Survey 5-Year Estimates subject tables. This information is not part of the census and thus constitutes an independent test.}
\begin{tabular*}{\hsize}{l @{\extracolsep{\fill}}rrrrrr}
\makecell[bl]{Census tract}&	\makecell[br]{\% pop. 25+\\w. bachelor's\\or higher}&	\makecell[br]{\% pop. below\\poverty level}&	\makecell[br]{Median household\\income (USD)}&	\makecell[br]{Median gross\\rent (USD)}&	\makecell[br]{Median house\\value (USD)}&	\makecell[br]{\% pop. 16+\\in labor force}\\
\midrule
\makecell[l]{743, Orange\\County, CA}&	3.6&    27.7&   55720&  1195&   329100& 69.0 \\
\makecell[l]{1042.01, Los\\Angeles\\County, CA}&	1.5&	19.0&	45089&	972&	291000&	69.3 \\
\makecell[l]{741.09, Orange\\County, CA}&	10.0&	14.2&	69688&	1401&	338600&	68.6 \\
\makecell[l]{747.01, Orange\\County, CA}&	3.1&	14.4&	58447&	1355&	309900&	67.8 \\
\makecell[l]{747.02, Orange\\County, CA}&	5.9&	23.0&	53169&	1117&	328200&	71.5 \\
\makecell[l]{23.05, Santa\\Barbara\\County, CA}&	3.4&	22.3&	52000&	1200&	205700&	62.8 \\
\makecell[l]{47.17, Ventura\\County, CA}&	4.0&	15.2&	65063&	1610&	288200&	72.6 \\
\makecell[l]{85.02, Hamilton\\County, OH}&	11.7&	84.4&	8878&	580&	70300&	53.1 \\
\makecell[l]{301, Lake\\County, IN}&	0.0&	79.5&	9504&	248&	180600&	73.8 \\
\makecell[l]{1143, Cuyahoga\\County, OH}&	0.8&	87.1&	8810&	298&	31000&	51.5 \\
\\
\\
\makecell[l]{307.05, Broward\\County, FL}&	21.9&	11.9&	26943&	985&	72500&	17.5 \\
\makecell[l]{77.47, Palm Beach\\County, FL}&	34.7&	12.1&	25482&	785&	70900&	14.1 \\
\makecell[l]{1551.01, Queens\\County, NY}&	60.0&	2.2&	70036&	1645&	410200&	25.2 \\
\makecell[l]{3511.02, Contra\\Costa County, CA}&	50.2&	4.8&	42827&	946&	207200&	18.7 \\
\makecell[l]{995.09, Orange\\County, CA}&	29.6&	8.6&	30188&	536&	192600&	15.0 \\
\makecell[l]{3511.01, Contra\\Costa County, CA}&	55.8&	1.4&	50385&	1312&	332500&	24.9 \\
\makecell[l]{77.46, Palm Beach\\County, FL}&	24.1&	8.8&	22813&	707&	47000&	20.0 \\
\makecell[l]{995.10, Orange\\County, CA}&	22.7&	17.3&	27223&	463&	153200&	15.3 \\
\makecell[l]{3511.03, Contra\\Costa County, CA}&	61.9&	4.7&	72625&	1538&	529900&	20.9 \\
\makecell[l]{405.13, Maricopa\\County, AZ}&	38.6&	1.2&	56471&	N/A&	211000&	7.1 \\
\bottomrule
\end{tabular*}

\end{table*}

We now inspect the eigenvector result (Fig.~2) more closely. In contrast to previous results from British cities, we do not see such a clear separation between middle class and deprived areas in the US. However there is a very strong separation between middle-class and highly affluent tracts, which are almost exclusively country clubs and certain holiday and retirement properties. This can be interpreted as an indication that the most significant social divide in the US is the separation of the affluent from the rest of society.  

A notable difference between the eigenvector-based index and the traditional poverty index is that the eigenvector highlights many Native American reservations as deprived areas \cite{saipe2010census}. Examples include parts of Wind River Reservation, Warm Springs Reservation, Hopi Reservation, and various pueblos in northern New Mexico.

The high spatial resolution of the eigenvector-index also allows to study differences that exist on a small scale. We find that these differences are extremely pronounced in Florida. The state harbors some of the most and least \cite{florida2019toward,sommeiller2018income} deprived tracts in nation, often in direct proximity. For comparison we also show the differences that exist for example in New York, between very affluent area east of Central park and borderline deprived areas in Harlem.  

An interesting feature can be seen along the route one between Washington and New York, where a thin \FloatBarrier \noindent stripe of increased deprivation follows the busy commuter route. Other notable areas of deprivation are seen along the lower Mississippi river, in California's Central Valley and South Texas. These are previously recognized areas of high deprivation \cite{wimberley2003us,calenviro2017OEHHA,bohn2011poverty,hotez2012texas}.

\section*{Summary and Conclusions}
In the present paper we have proposed manifold-learning with diffusion maps as a method to generate a deprivation index from census data. 
One advantage of this method is that it can utilize large amount of data contained in the census. The diffusion map uses this data to assign new variables to census tracts and thus reveal similarities in living conditions. The observed patterns of similarity include some that are induced by specific external circumstances such as enrollment in the military, but also reveal the effect of deprivation. Our results reveal well-known large scale deprivation in certain areas, but also highlight the feature at the local level. 

What makes diffusion maps particularly attractive for constructing deprivation indices is that they are strongly resistant to  manipulation and don't introduce biases beyond those inherent in the source data. The only choices to make in this method are the construction of the similarity matrix (here the inverse of Euclidean distance, an intuitive non-parametric choice) and the way thresholding is done (here we kept the top-9 strongest links, again an intuitive choice). We found the results to be very robust to sensible variations and outperformed other methods (PCA, k-means, see Appendix). Perhaps more importantly, even armed with a detailed understanding of the diffusion map it is not possible to make choices in such a way as to create a specific result, which greatly limits the potential to manipulate the procedure for ideological reasons. 

Even the diffusion map does not remove human intuition entirely from the data analysis process as interpretation is still needed to interpret the meaning of the eigenvectors. These interpretations should be regarded as hypotheses that are then confirmed or rejected through additional tests. Here we performed such a test by correlating the respective eigenvector with existing deprivation indices in the area of Los Angeles where a detailed index was available.

Beyond the first four eigenvectors the diffusion map reveals other eigenvectors that are of lesser importance overall but may still hold interesting information that yields deeper insights. We hope that the exploration of these eigenvectors will provide additional understanding of the social geography of the US in the future.

\section*{Acknowledgements}
TG thanks the Ministry of Science and Culture of Lower Saxony and the Volkswagen Foundation (grant ZN3285) for support.

\section*{References}
\bibliographystyle{unsrtnat}

\renewcommand{\bibsection}{}
\bibliography{references}

\newpage
\section*{Appendix}
\subsection*{US census data}
The decennial census has been conducted in years ending in zero since 1790 \cite{censusgovwebsite}. The latest completed census is the 2010 census, and the data is compiled from the questions asked of all people and about every housing unit. Population items include sex, age, race, Hispanic or Latino origin, household relationship, household type, household size, family type, family size, and group quarters. Housing items include occupancy status, vacancy status, and tenure \cite{uscensustechnical}.

The data is available at several geographic levels including states, counties, and census tracts, which are small, relatively permanent statistical subdivisions of a county or county equivalent and generally have a population size between 1200 and 8000 people with an optimum size of 4000 people \cite{uscensustechnical,us2010census}. Amount of available data is subject to the choice of geographical summary level and generally more data is available for larger geographical summary levels.

\subsection*{Data preprocessing}
In this analysis we consider the census tract summary level. There are 73057 census tracts among the 50 states and Washington, D.C. in the 2010 census, and over 8500 data columns are available at this summary level. Of these, we remove columns for which either the feature itself or the surveyed universe incorporate racial information. The resulting dataset consists of 1385 entries for each census tract.

A number of census tracts have zero population. These are typically tracts that cover uninhabited terrains such as bodies of water. These tracts are dismissed for the rest of the analysis. There are a few tracts for which a range of entries are missing. These tracts are also removed at this stage and will be processed later. See below for a description of how these tracts are treated.

Census tracts were intended to have the same or very close populations so that reasonable comparisons can be made. Even though there is noticeable nonhomogeneity in census tract populations, by simply dividing count entries of a census tract by its population (which is itself an entry) we can shape the desirable intensive variables for comparison. The resulting dataset is stored as an $M \times N$ matrix $\bf X$ where $M$ is the number of remaining tracts and $N$ is the number of normalized census variables.

\subsection*{Diffusion maps}
To implement diffusion maps we begin by standardizing all the columns to ensure that various variables will be of the same scale and with all origins set at the mean. This is done through enforcing a mean of 0 and a standard deviation of 1 for all variables using

\begin{align}
    \hat{X}_{ij} = \frac{X_{ij}-\mu_j}{\sigma_j},
\end{align}
where $\mu_j$ and $\sigma_j$ are respectively the mean and standard deviation of column $j$ of matrix $\bf X$.

Each of the $M$ rows can be thought of as a point in the $N$-dimensional space where entries are coordinates that specify the corresponding census tract. We then define an $M \times M$ distance matrix $\bf D$ where

\begin{align}
    D_{ij} = \sqrt{\sum_{k}{(\hat{X}_{ik}-\hat{X}_{jk})^2}}
\end{align}
is the Euclidean distance between census tracts $i$ and $j$.

Next, we form the similarity matrix $\bf C$ such that

\begin{align}
C_{ij}=
\begin{cases}
      k(D_{ij}) & i \neq j \\
      0 & i = j
\end{cases} 
\end{align}
where $k$ is an appropriate kernel. The chosen kernel must be such that $\bf C$ is symmetric and $C_{ij}\ge 0$. Here we use $k(d)= 1/d$. The higher value of $C_{ij}$ now indicates higher similarity between census tracts $i$ and $j$.

Since Euclidean distance is typically a local metric in the feature space, it is often necessary to disregard long distances beyond the local neighborhood of a census tract in the space of data. Equivalently, we can "threshold" the similarity matrix $\bf C$ to only keep the scores between each tract and its closest neighbors and set other entries to zero. There are various ways this can be done; for instance, a minimum similarity score can be assumed below which the similarities are dismissed.

Alternatively, we use a heuristic procedure where we keep the similarities between a census tract and a fixed number of its neighbors with highest similarity scores and set the remaining entries to zero. In case there are non-symmetric instances (i.e.~the modified similarity matrix is not symmetric), the asymmetric zero entries are updated to their corresponding non-zero values. $\bf C$ can now be regarded as the weighted adjacency matrix of a network. We use a threshold of 9 neighbors for each census tract, but as long as the threshold is small enough and the network remains connected the results would not be much affected. Another benefit of thresholding is that it allows for much higher computational efficiency as the modified similarity matrix will be sparse.

We now construct the network's random-walk normalized Laplacian matrix $\bf L$ such that

\begin{align}
L_{ij}=
\begin{cases}
      -\frac{C_{ij}}{\sum_{k}C_{ik}} & i \neq j \\
      1 & i = j
\end{cases} 
\end{align}
This matrix is related to the transition matrix of a random walk in a network and describes diffusion processes on the network nodes \cite{coifman2005geometric}.

We can now obtain an embedding of the census tracts on a low dimensional manifold through finding eigenvalues and eigenvectors of $\bf L$. $\bf L$ is positive semi-definite and therefore is guaranteed to have real non-negative eigenvalues. The number of zero eigenvalues is identical to the number of components in the network \cite{chung1997spectral} and hence we expect to find exactly one zero eigenvalue. The eigenvector corresponding to this eigenvalue does not carry any information and is disregarded. 

The most relevant eigenvalues for our purpose are the smallest non-zero eigenvalues as their corresponding eigenvectors place the census tracts appropriately along the main directions of the low dimensional manifold that captures the most essential variations in data. Each eigenvector assigns a value to each census tract that can be used to assess where that tract is located along the corresponding direction on the manifold.

\subsection*{Tracts with partially missing data}
To assign appropriate values to census tracts with partially missing data for each eigenvector, we calculate the Euclidean distances between each of these tracts and all other tracts based on commonly available standardized data columns. For each eigenvector, the value assigned to such a census tract will be the same value as its closest neighbor \cite{barter2019manifold}.

\subsection*{Analysis with PCA and \textit{k}-means clustering}
We also analyzed the census dataset with PCA \cite{pearson1901liii,jolliffe2011principal,moon2000mathematical} and \textit{k}-means clustering \cite{macqueen1967} with the same standardization. The motivation behind using the \textit{k}-means clustering is that social deprivation might in fact be an explaining factor for the variation in the dataset to the extent that clusters broadly capture areas of the same level of deprivation. For clustering we used 20 clusters.

The first PCA variable picks up traces of deprivation, while the result from \textit{k}-means clustering does not align with deprivation when considering the unsupervised clusters. It is, however, possible to hypothesize that the clusters indeed correspond to various levels of deprivation and then try to, for instance, maximize the rank correlation between cluster IDs of census tracts and a measure of deprivation, but that means pinning down a specific choice of cluster IDs from among $20!$ possible configurations in this case such that the new index will be optimally close to an established index. A trivial example for this would be when the number of clusters equal the number of census tracts, in which case doing this optimization would simply result in ordering of the census tracts based on the existing index. Further, even a successful clustering result means that the deprivation can only be more broadly classified and smaller differences would vanish in this index.

\begin{figure}
\centering
\begin{subfigure}{\textwidth}
  \centering
  \includegraphics[width=\textwidth]{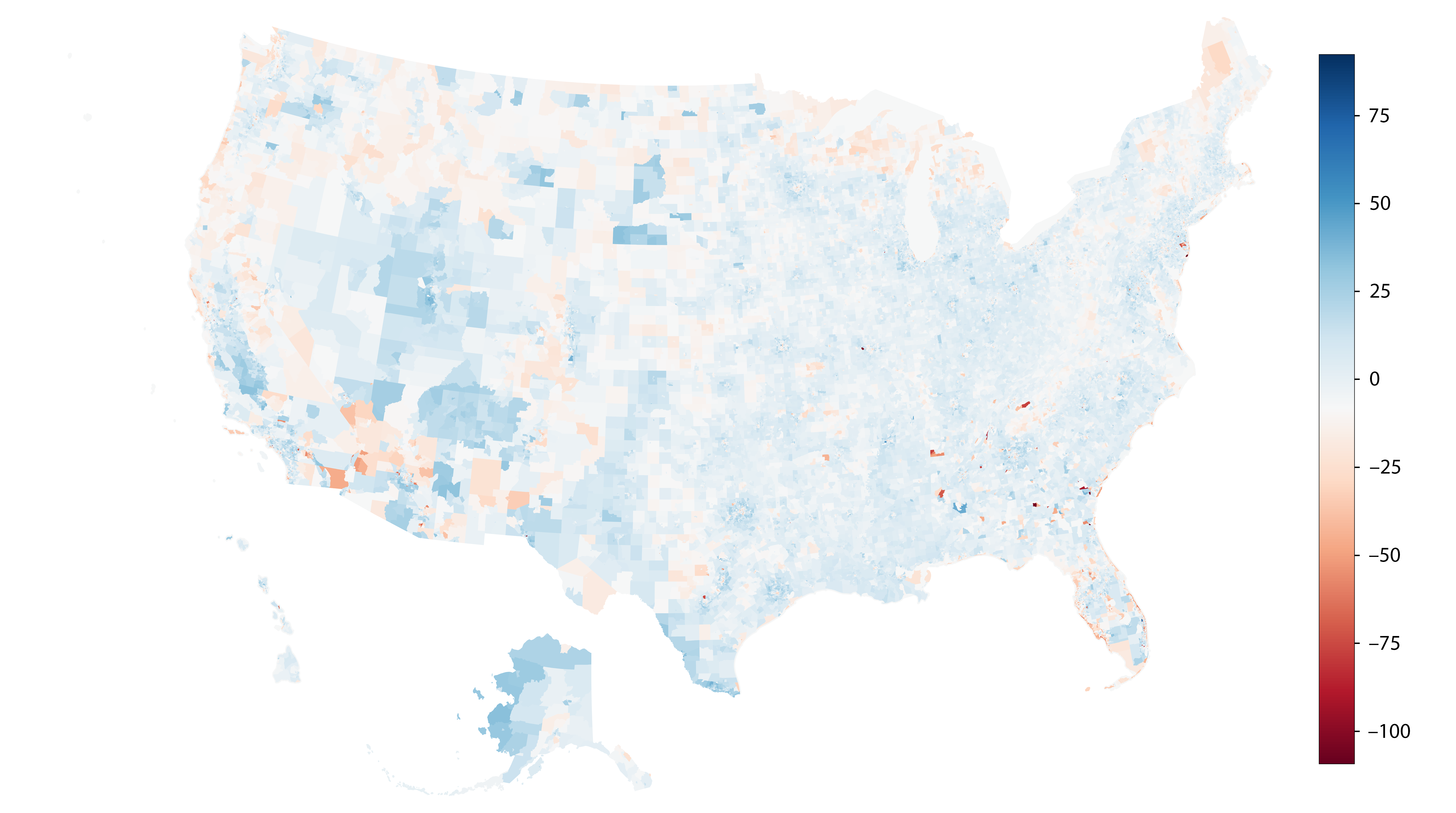}
    \caption{}
\end{subfigure}
\begin{subfigure}{\textwidth}
  \centering
  \includegraphics[width=\textwidth]{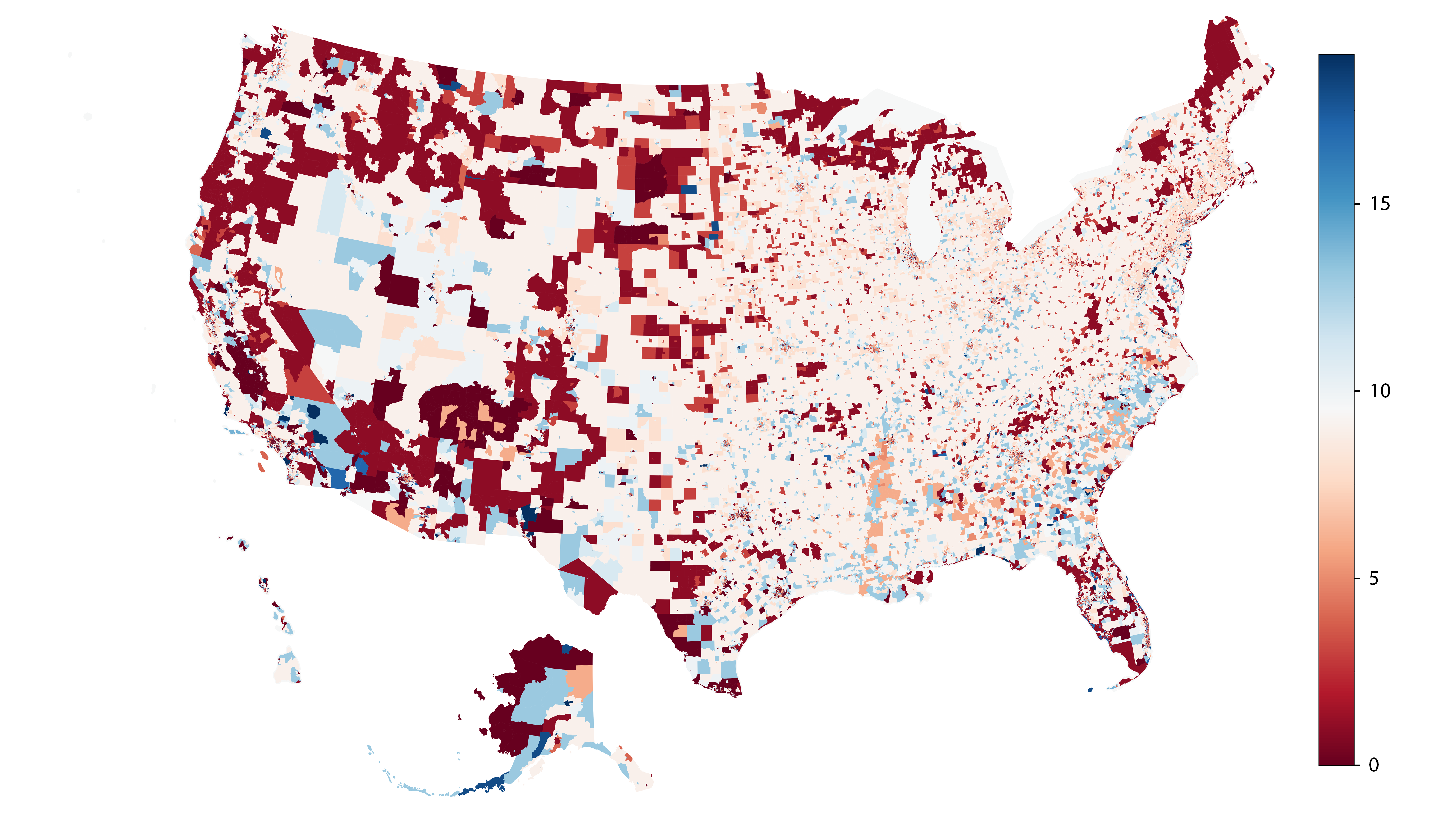}
    \caption{}
\end{subfigure}
\caption{Results from first PCA feature (a) and \textit{k}-means clustering with 20 clusters with labels 0 through 19 (b). Deprivation appears to partially explain the variations in census tract assigned values in PCA. For \textit{k}-means clustering, the output numerical cluster ordering may or may not be optimal in aligning census tracts with deprivation without additional evaluation. Also, smaller distinctions between various census tracts will disappear as a result of clustering.}
\end{figure}

\end{document}